\documentclass[aps,pra, twocolumn,a4paper,showpacs]{revtex4}
\usepackage{amsmath}
\usepackage{amssymb}
\usepackage{bm}
\usepackage{epsfig}
\usepackage{graphicx}

\def\BE {\begin{equation}}
\def\EE {\end{equation}}
\def\BA {\begin{array}}
\def\EA {\end{array}}

\begin{document}

\title{Efficient and long-lived field-free orientation of molecules by a
single hybrid short pulse}

\author{D.~Daems $^{1, 2}$, S.~Gu\'erin $^{1}$, D. Sugny,$^{1}$ and H. R.~Jauslin$%
^{1}$}
\email{ddaems@ulb.ac.be, sguerin@u-bourgogne.fr}
\affiliation{$^{1}$Laboratoire de Physique de l'Universit\'e de Bourgogne, UMR CNRS 5027,
BP 47870, 21078 Dijon, France\\
$^{2}$Center for Nonlinear Phenomena and Complex Systems, Universit\'e Libre
de Bruxelles, 1050 Brussels, Belgium}

\begin{abstract}
We show that a combination of a half-cycle pulse and a short
nonresonant laser pulse produces a strongly enhanced postpulse
orientation. Robust transients that display both
efficient and long-lived orientation are obtained. The mechanism is
analyzed in terms of optimal oriented target states in finite
Hilbert subspaces and shows that hybrid pulses can prove useful for other control issues.
\end{abstract}

\pacs{33.80.-b, 32.80.Lg, 42.50.Hz}
\maketitle

Laser controlled processes such as molecular alignment and
orientation are challenging issues that have received considerable
attention both theoretically and experimentally
\cite{Seideman_review}. Whereas strong nonresonant adiabatic
pulses can exhibit efficient alignment and orientation only when
the pulse is on \cite{Friedrich,Vrakking,PRL2002}, linear polar
molecules can be oriented under field-free
conditions after the extinction of a short half-cycle pulse (HCP) \cite%
{Dion01,Machholm}. Its highly asymmetrical temporal shape imparts
a sudden momentum kick through the permanent dipole moment of the
molecule, which orients it. This extends the use of a permanent
static field (combined with pulsed nonresonant laser fields)
\cite{Cai,Sakai}. Similarly to the alignment process by a
nonresonant short pulse \cite{Rabitz} (measured by $\langle \cos
^{2}\theta \rangle $ with $\theta $ the angle between the axis of
the molecule and the polarisation direction of the laser field),
the orientation by a HCP (measured by $\langle \cos \theta \rangle
)$ increases as a function of the field amplitude until it reaches
a saturation at $\langle \cos \theta \rangle \approx 0.75$, which
corresponds to an angle $\cos ^{-1}\langle \cos \theta \rangle
\approx 41^{\circ }$. Overcoming this saturation has been
theoretically proved with the use of trains of laser pulses (HCP
kicks) in the case of alignment \cite{Rabitz} (orientation \cite%
{reaching}). For applications it is of importance to reach an
efficient orientation. Another crucial point, that has received
less attention so far \cite{reaching, Ortigoso}, is the duration during which the
orientation is above a given threshold, which one would like to
keep as long as possible. An important step made in
\cite{reaching} was to establish a priori the two oriented target
states (of opposite direction) in a given finite subspace
generated by the lowest rotational states. These target states are
optimal in the sense that they lead respectively to the maximum
and minimum values of $\langle \cos \theta \rangle$ in this given
subspace. The choice of a suitable small dimension of the subspace
allows one to generate an oriented target state of relatively long
duration. The identification of such an optimal target state opens
in particular the possibility to use standard optimization
procedures (see e.g. \cite{AtabekDion}).

One of the main challenges consists now in reaching such an
optimal target state in a subspace of low dimension,
characterizing a long-lived and efficient orientation, by a simple
external field in a robust way and to ensure the persistence of
this effect with respect to thermal averaging of finite
temperature. We propose in this Letter a process that possesses
such properties. By superimposing a pump laser field to a
half-cycle pulse, we show that the maximal orientation reached
after the pulse is significantly beyond the one induced by an HCP,
and that it displays furthermore a larger duration. We obtain in
particular the saturation $\langle \cos \theta \rangle \approx
0.89$, which corresponds to the angle $\cos ^{-1}\langle \cos
\theta \rangle \approx 27^{\circ }$. This efficient and long-lived
orientation is obtained by adjusting only two parameters: the
amplitudes of the laser and HCP fields. We obtain robust regions
of the parameters generating this orientation.

We show that this process allows one to approach an optimal target
state in one step. Under the action of a single
hybrid pulse, the number of rotational states that are
significantly populated remains finite and controllable. This
generates a finite dimensional subspace in which an optimal target
state can be constructed. When the dimension of this subspace
increases, the associated optimal state yields a higher
orientation efficiency while its duration decreases. By choosing
appropriate intensities of the pump laser field and of the
half-cycle pulse, we can both select and reach the target state
with the desired efficiency and duration.

We consider a linear molecule in its ground vibronic state described in the
3D rigid rotor approximation. The effective Hamiltonian including its
interaction with a HCP simultaneously combined with a pump laser field of
respective amplitudes $\mathcal{E}_{\text{HCP}}(t)$ and $\mathcal{E}_{\text{L}}(t)$
is given by
\begin{equation}
H_{\text{eff}}(t)=BJ^{2}- a_{\text{HCP}}(t)\cos \theta -a_{\text{L}}(t)\cos
^{2}\theta  ,  \label{model}
\end{equation}%
where $B$ is the rotational constant, $a_{\text{HCP}}=\mu _{0}\,\mathcal{E}_{\text{HCP}}$ with $\mu _{0}$
the permanent dipole moment, and $a_{\text{L}}=\Delta \alpha \,\mathcal{E}_{%
\text{L}}^{2}/4$ with $\Delta \alpha $ the polarisability
anisotropy. Note that $\Delta \alpha $ is positive for linear
molecules, which gives positive values for $a_{\text{L}}$ whereas
the sign of $a_{\text{HCP}}$ is determined by the sign of
the HCP amplitude $\mathcal{E}_{\text{HCP}}$.
The dynamics of the system is readily determined  with the help of the propagator in the impulsive regime, where the
duration $\tau$ of the pulse is much smaller than
the rotational period $\tau_{\text{rot}}=\pi\hbar/B$ \cite{Henriksen}.
For the process we suggest here, in the dimensionless time
$s=t/\tau_{\rm {rot}}$ whose origin coincides
with the extinction time of the pulse, the propagator reads
$
U(s,0)=e^{-i \pi J^{2}s}e^{iA_{\text{HCP}}\cos \theta }e^{iA_{%
\text{L}}\cos ^{2}\theta }.
$
The parameters $A_{\text{HCP}}=\frac{1}{\hbar }\int dt\ a_{\text{HCP}}(t)$ and $A_{%
\text{L}}=\frac{1}{\hbar }\int dt\ a_{\text{L}}(t)$ are
respectively the total  dimensionless 
areas of the HCP amplitude and of the laser intensity.

We first consider the case of a cold
molecule whose state after the pulse reads
$|\phi(s)\rangle=U(s,0)|j=0\rangle$ where $|j\rangle$ stands for the spherical harmonics $Y_j^m$ with $m=0$.
The orientation is measured by the expectation value $\langle \cos
\theta \rangle (s)=\langle \phi(s)|\cos\theta|\phi(s) \rangle $.
It is well-known that it is a periodic function (of unit period), which exhibits peaked revivals corresponding to
molecular orientation along the field direction
\cite{Dion01,Machholm}.

\begin{figure}[h]
\centerline{\includegraphics[scale=0.63]{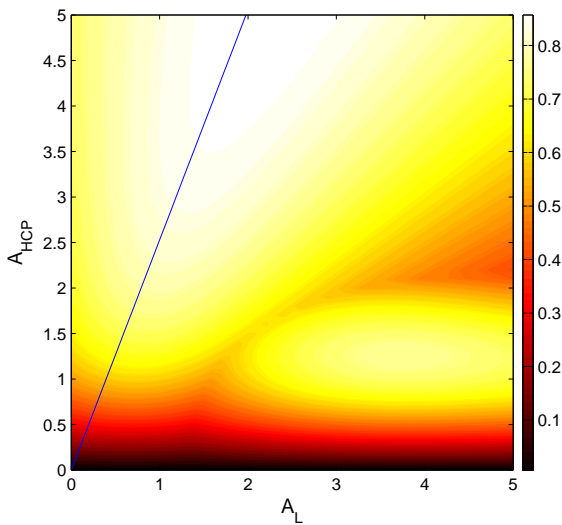}}
\caption{(Color online) Numerical contour plot of
$\max_s|\langle\cos\theta\rangle|$ as a function of the total
(dimensionless) areas $A_{\text{L}}$ and $A_{\text{HCP}}$. The
straight line  $A_{\text{HCP}}=2.5\,A_{\text{L}}$
approximately indicates the maximum for given
 $A_{\text{HCP}}$.} \label{contour1}
\end{figure}

 The maximal field-free orientation is displayed in Fig. \ref{contour1} by plotting
the maximum of $|\langle\cos\theta\rangle|$ reached over a period as a function of the total areas
$A_{\text{L}}$ and $A_{\text{HCP}}$ at zero rotational
temperature, calculated with the above propagator. The
case of a single HCP coincides with the ordinate axis. In the
absence of the laser pulse it is seen that the maximal orientation
saturates to a value of $\langle\cos \theta\rangle \approx 0.75$.
In the presence of a simultaneous pump laser pulse, one observes a
wide two-dimensional plateau as well as an island which are both
associated with an orientation much higher than the saturation
limit of a single high intensity HCP. The plateau is relatively flat in a large
two-dimensional region centered approximately around the line
$A_{\text{HCP}}= 2.5 \, A_{\text{L}}$, implying that one
can robustly reach a high efficiency for the orientation with a
moderate HCP intensity.
The island observed around
$A_{\text{HCP}}=1.25$ and $A_{\text{L}}=3.7$ indicates that it is
also possible to overcome the above saturation by combining an HCP of intensity slightly above
unity  with a laser pulse of high
intensity.

The direction of the orientation can be chosen by the sign of the
amplitude of the HCP. 
Expressing
the observable in terms of the projections $c_j=\langle
j|\phi(0)\rangle$ of the wave function right after the pulse onto
the rotational states $|j\rangle$ leads to
$
\left\langle \cos\theta\right\rangle (s) =\frac{1}{2}\sum c_j^\star c_{j+1} e^{-2i \pi (j+1) s}+ {\text c. \text c.} .$
The coefficients $c_j$ can be calculated for the  above propagator
 in the approximation $\langle
j|\cos\theta|j\pm1\rangle\simeq1/2$ which is more accurate for
$j\gg 0$ (one has $\langle 0|\cos\theta|1\rangle\approx0.58$,
$\langle 1|\cos\theta|2\rangle\approx0.52$, $\langle
2|\cos\theta|3\rangle\approx0.51$, $\cdots$). This shows
that the sign of each product $c_j^\star c_{j+1}$, and hence of
$\left\langle \cos\theta\right\rangle$, changes with the sign of
$A_{\text{HCP}}$. For positive  $A_{\text{HCP}}$, the expectation value corresponding to $\max_s|\langle\cos\theta\rangle|$ is positive on the island and negative on the plateau. 
We show below that the high orientation efficiency obtained
in the region along the line $A_{\text{HCP}}=
2.5\, A_{\text{L}}$ has the remarkable property to approach very
closely an optimal state as defined in \cite{reaching}, which
combines orientation of high efficiency and of long duration.

\begin{figure}[h]
\centerline{\includegraphics[scale=0.63]{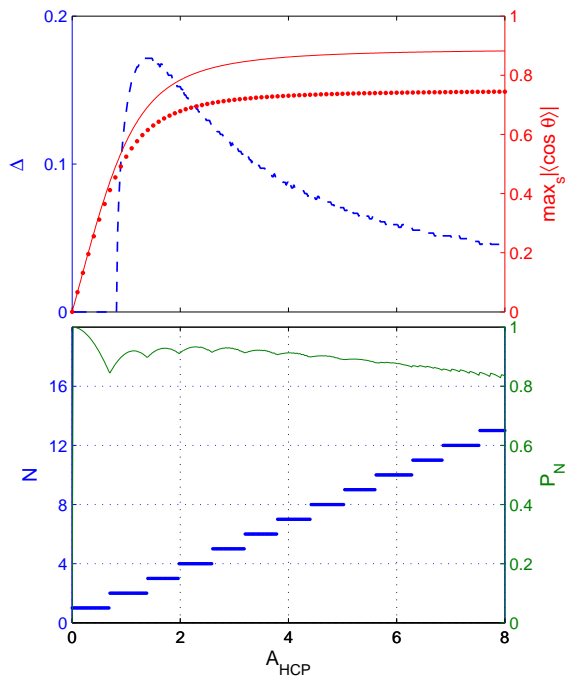}}
\caption{As a function of $A_{\rm{HCP}}$, for $A_{\rm{L}}=A_{\rm{HCP}}/2.5$, upper
panel: (i) $\max_s|\langle\cos\theta\rangle|$  (solid line, right axis);(ii) for comparison, same quantity with $A_{\rm{L}}=0$ (dotted line); (iii) duration $\Delta$ for which
$|\langle\cos\theta\rangle|>0.5$ (dashed line, left axis);
Lower panel: square
modulus
$P_N=|\langle\chi_-^{(N)}|\phi(s_{\max})\rangle|^2$  of the projection of the state at the time  giving the
maximum  of $|\langle\cos\theta\rangle|$  on the optimal state $|\chi_-^{(N)}\rangle$ (solid line, right
axis);  dimension $N$ of the corresponding
 subspace (step function, left
axis).} \label{fig2}
\end{figure}

\begin{figure}[h]
\centerline{\includegraphics[scale=0.63]{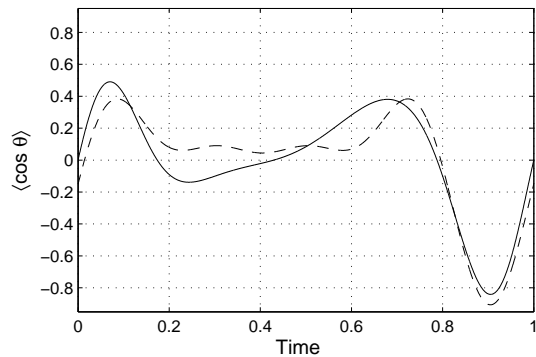}}
\caption{Orientation as a function of dimensionless time for (i)
$A_{\text{HCP}}=3$ and $A_{\text{L}}=3/2.5$ (solid line) and (ii)
the optimal state $\chi_-^{(5)}$ with a time translation (dashed line).} \label{timeline}
\end{figure}

The upper frame of Fig. \ref{fig2} compares the maximal value of
$|\langle\cos\theta\rangle|$ as a function of $A_{\text{HCP}}$ (i)
without the laser, and (ii)
along the straight line $A_{\text{HCP}%
}=2.5 \, A_{\text{L}}$ seen in Fig. 1. This shows that the saturation
of $\langle \cos \theta \rangle \approx 0.75$ obtained
with the HCP alone is significantly overcome (up to $\langle \cos
\theta \rangle \approx 0.89$) when the HCP is associated with the
laser of appropriate area. The upper frame of Fig. \ref{fig2} also
illustrates the duration of the orientation 
defined as the time during which $|\langle \cos \theta
\rangle|\ge0.5$ (see also
Fig. \ref{timeline}). It is seen that the duration of the revival
can be as large as about 18$\%$ of the rotational period. The properties displayed in Fig. 2 are robust with
respect to the parameters $A_{\text{HCP}}$ and $A_{\text{L}}$
which need not be  in a strict 2.5 ratio.

The efficiency of the obtained oriented state and its  large
duration are explained in terms of an optimal state as defined in
\cite{reaching}. We recall that the optimal states correspond to
the two states that respectively minimize and maximize the
projection  of $\cos\theta$ in the finite subspace ${\cal H}_N$
spanned by the $N$ lowest rotational states
$|0\rangle,|1\rangle\,\cdots,|N-1\rangle$, namely
$\cos^{(N)}\theta=\Pi_N\cos\theta \Pi_N$ with the projector
$\Pi_N=\sum_{j=0}^{N-1}|j\rangle\langle j|$. Considering a finite
subspace yields an operator  that has a discrete spectrum, whose eigenvectors are readily  calculated and for which the 
duration of the orientation provided by these states can be
computed. Furthermore, the controllability of the system can be completely analyzed \cite{Ramakrishna,Schirmer}. For a given dimension $N$, the two optimal states are
the eigenvectors associated respectively with the smallest and
the largest eigenvalues of $\cos^{(N)}\theta$. In the
approximation $\langle j|\cos\theta|j\pm1\rangle\simeq 1/2$  one
obtains
\begin{equation}
|\chi_\pm^{(N)}\rangle\simeq \sqrt{\frac{2}{N+1}}\sum_{j=0}^{N-1}(\pm 1)^{j+1}
\sin\left(\pi\frac{j+1}{N+1}\right)|j\rangle, \label{chi}
\end{equation}
giving the approximate optimal orientation
\begin{equation}
\langle\chi_{\pm}^{(N)}|\cos^{(N)}\theta|\chi_{\pm}^{(N)}\rangle\simeq
\pm \cos\left(\frac{\pi}{N+1}\right).
\end{equation}
The (relative) duration $\Delta$ of the orientation  is defined as the time during which $|\langle \cos \theta
\rangle|\ge \gamma$ for the revival of maximum efficiency, with $\gamma$ arbitrarily chosen as $1/2$. 
We can determine the duration $\Delta_N$ for the state $|\chi_\pm^{(N)}\rangle$  by summing the  above expression for $\left\langle \cos\theta\right\rangle (s)$ and expanding the result to second order around its extremum, obtaining
\begin{equation}
\Delta_N  \approx  \frac{2}{\pi} \sqrt{\frac{1}{\Gamma_N}\left[1-\gamma/{\cos\left(\frac{\pi}{N+1}\right)}\right]},
\end{equation}
where $\Gamma_N=\alpha (N+1)^2-(N+1)$ with $\alpha=2/3-1/\pi^2$.
The shape of $\Delta_N$ as a function of $N$ is similar to the dashed curve on the upper panel of Fig.~2, independently of the specific value of $\gamma$.
In particular, the  decrease of this duration for large $N$ is due to the factor $\Gamma_N$.

Finding a process that drives
the system to an optimal state $\chi_\pm^{(N)}$   guarantees an efficient orientation together with a
 large duration if the dimension $N$ of the subspace ${\cal
H}_N$ generated by the dynamics is relatively low \cite{reaching}.
The lower frame of Fig. \ref{fig2} shows that the HCP-laser
combination of appropriate areas leads in a single step to a wave
function that is remarkably close  to the optimal  state
$|\chi_{-}^{(N)}\rangle$ (more than $90\%$ for
$1.5<A_{\text{HCP}}<5$). Figure \ref{fig2} also indicates how the
dimension $N$ of the embedding subspace can be chosen by the value
of $A_{\text{HCP}}$ with $A_{\text{L}}=A_{\text{HCP}}/2.5$. Notice the  linear character  of this necessarily stepwise function.
In the island region of Fig. (\ref{contour1}), the dynamics also
generates a state close to an optimal one:
$|\langle\chi_+^{(6)}\vert\phi(s_{\max})\rangle|^2\approx 0.86$
for $A_{\text{HCP}}=1.25$ and $A_{\text{L}}=3.7$.
For comparison, we note that the same optimal states are reached in \cite{reaching} 
 by a different process involving 15 short HCP kicks sent
at  specific times  and with a
low amplitude in order to remain in a given subspace.

Figure 3 shows an example of orientation, measured by $\langle \cos\theta \rangle$, as a function of time for a point on the straight line of Fig. 1 ($A_{\rm{HCP}}=3)$.
The result is close to that given by the optimal state  $\chi_-^{(5)}$
whose minimum is taken at the minimum of $\langle\cos
\theta\rangle$ generated by the hybrid pulse. We obtain in this
case $|\langle\chi_-^{(5)}\vert\phi(s_{\max})\rangle|^2\approx
0.93$ (with less than $2\%$ of the total
population outside the subspace ${\cal H}_5$.) The maximum
orientation (in absolute value) $|\langle\cos
\theta\rangle|\approx0.9$ occurs at $s\approx 0.9$. One can
observe a relatively large duration of the orientation. In contrast to
the case of a sole HCP, the presence of the laser pulse of
appropriate area ($A_{\text{L}}=A_{\text{HCP}}/2.5$)  allows us to
obtain, immediately after the pulse, projections $c_j$ on the
rotational states whose moduli are very close to the moduli of the
corresponding components of both optimal states. The phases of these projections just
after the pulse generally differ from those of the components of
the optimal states, but are brought by the free evolution closer
to those of one or the other optimal state. In the case of the
plateau region, the set of phases after the pulse leads to the
state $\chi_-^{(N)}$ for positive   $A_{\text{HCP}}$ and yields
thus a minimal value for $\langle\cos \theta\rangle$.
This analysis extends to the island region where revivals of
opposite sign are observed. As discussed above, by changing the
sign of $A_{\text{HCP}}$ while keeping $A_{\text{L}}$ fixed, one
obtains orientation revivals of the same absolute values but
opposite sign. The direction of the orientation can thus be
controlled  by the sign of the HCP pulse.

\begin{figure}[h]
\centerline{\includegraphics[scale=0.63]{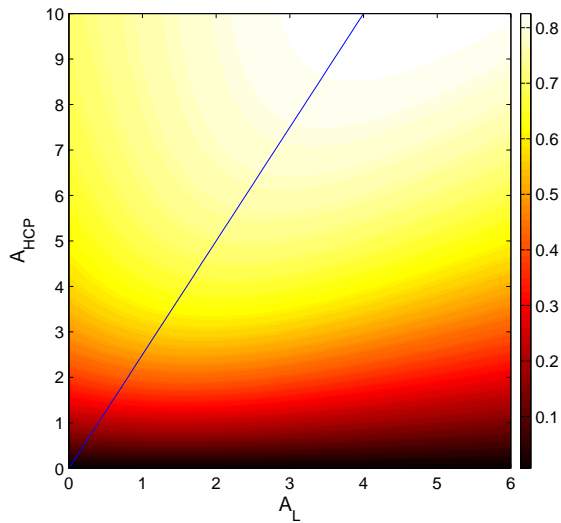}}
\caption{Same as Fig. \ref{contour1} for the
dimensionless temperature $\tilde T=5$.} \label{contourT}
\end{figure}

Considering the effect of temperature amounts to statistically
average over the solutions of the Schr\"odinger equation with
different initial conditions $|j,m\rangle$ weighed by a Boltzmann factor.   Figure
\ref{contourT} shows the maximal  orientation, measured by the appropriate expectation value of $\cos \theta$,  as a function of the field parameters for a dimensionless temperature $\widetilde T \equiv kT/B=5$ (which corresponds to $T\approx5$ K for
the LiCl molecule). Notice that the island disappears while the region around the
straight line $A_{\text{L}}=A_{\text{HCP}}/2.5$ persists. The
efficient and  long-lived orientation revivals are therefore robust with respect to  thermal 
averaging and to the field parameters. The efficiency is lower than at $T=0$K for the
same field amplitudes, but one can recover the same value by
increasing the amplitudes along the straight line.

In conclusion, we have shown that a combination of a half-cycle
pulse and a short nonresonant laser pulse of appropriate
amplitudes leads  to efficient and long-lived revivals of
orientation beyond the known saturation. Furthermore, this is achieved in a controllable manner since the desired target state can be chosen in a set of optimal target states defined in  Hilbert
subspaces of low dimension and be reached with a projection larger than 90$\%$ by a single hybrid pulse. As an illustration,  the ground state of a KCl molecule with
rotational constant $B\approx0.13$ cm$^{-1}$
($\tau_{\text{rot}}\approx128$ ps) and dipole moment
$\mu_0\approx10.3$ D gives $A_{\text{HCP}}\approx3$ for a pulse
duration of 2 ps and a HCP amplitude of 100 kV/cm. To be 
on the optimal  line of Figs. \ref{contour1} or 4 requires
$A_{\text{L}}\approx1.2$ which corresponds to a peak intensity 
$I\approx 10^{11}$ W/cm$^2$ for the laser field. These parameters lead to $\max_s|\langle\cos
\theta\rangle|\approx0.85$ an a duration of approximately
$1/8^{\text{th}}$ of the rotational period for a cold molecule
(see Fig. \ref{timeline}), and to $\max_s|\langle\cos
\theta\rangle|\approx0.73$ and a duration of approximately
$1/20^{\text{th}}$ of the rotational period for $T=5$ K.
The interest of hybrid pulses is not limited to molecular orientation but extends to  optimization issues of a large class of systems where symmetries need to be broken or selectively addressed (e. g. the control of tunneling).
The central element consists in using an external field that plays individually  on couplings of different 
symmetries.
In order to drive the dynamics even closer to an optimal 
target state, standard optimization algorithms can be used for trains of
these hybrid pulses (with for instance the delays and/or the
relative amplitudes between the kicks, or even a delay between the HCP and laser pulses) and should
require only a low number of hybrid pulses since the first step  already brings the system very close to
the target state. 

This research was supported in part by the \textit{Conseil R\'{e}gional de Bourgogne} and the \textit{Action
Concert\'{e}e Incitative Photonique} from the French Ministry of
Research.

\end{document}